\documentclass[3p,11pt]{elsarticle}

\usepackage{amssymb,amsmath,xcolor}

\usepackage{hyperref}

\makeatletter
\def\ps@pprintTitle{%
 \let\@oddhead\@empty
 \let\@evenhead\@empty
 \def\@oddfoot{}%
 \let\@evenfoot\@oddfoot}
\makeatother

\def\zeti{{\tilde \zeta}}
\def\r3{\sqrt{3}}
\def\ep{\epsilon}
\def\tzeta{{\tilde \zeta}}
\def\cO{{\cal O}}
\newcommand{\ra}{\rightarrow}

\newcommand{\be}{\begin{equation}}
\newcommand{\ee}{\end{equation}}
\def\beq#1\eeq{\begin{align}#1\end{align}}
\newcommand{\nn}{\nonumber}

\newcommand{\cN}{{\cal N}}
\biboptions{sort&compress}
\begin{document}

\begin{frontmatter}



\title{A perturbative study of holographic mABJM theory}


\author[1,2]{Nakwoo Kim\corref{a1}}
\ead{nkim@khu.ac.kr}
\author[1,3]{Se-Jin Kim}
\ead{sejin.kim@cern.ch}

\cortext[a1]{Corresponding author}

\address[1]{Department of Physics and Research Institute of Basic Science,
	Kyung Hee University,\\ Seoul 02447, Republic of Korea}
	
\address[2]{School of Physics, Korea Institute for Advanced Study, Seoul 02445, Republic of Korea}

\address[3]{Theoretical Physics Department, CERN, CH-1211 Geneva 23, Switzerland}

\begin{abstract}
Recently the calculation of holographic free energy for mass-deformed ABJM model (mABJM) with ${\cal N}=2$ supersymmetry and $SU(3)\times U(1)$ global symmetry was tackled by Bobev et al. \cite{Bobev:2018wbt}. We solve the associated BPS equations, requiring IR regularity, using a perturbative method proposed by one of us in \cite{Kim:2019feb}. In particular, we provide an analytic proof of a crucial conjecture made in \cite{Bobev:2018wbt} based on numerical solutions:  that the R-charge values of three chiral multiplets in mABJM should be independent of the IR values of a hypermultiplet scalar, which is holographically dual to the superpotential mass term.
\end{abstract}


\end{frontmatter}


\section{Introduction and Summary}
\label{sec:intro}
Supersymmetric localization techniques enable us to compute some BPS quantities exactly for supersymmetric gauge field theories in appropriately chosen backgrounds. See {\it e.g.} \cite{Pestun:2007rz,Kapustin:2009kz,Jafferis:2010un,Hama:2010av,Kallen:2012va} and also \cite{Pestun:2016zxk} for a review and more complete list of references. Such analytic results can be of course used to check various string dualities. In particular, we are here interested in verifying AdS/CFT correspondence \cite{Maldacena:1997re} with a broken conformal invariance by relevant deformations.
More specifically, on gauge field theory side one can compute quantities such as the partition function and Wilson loops when the field theory is put on the sphere. Then on the gravity side, one takes the relevant 10/11 dimensional supergravity in Euclidean signature and solve the BPS equations when the modes whose dual operators we consider are turned on. We look for solutions which asymptote to Euclidean AdS, {\it i.e.} hyperbolic space  in the UV (when the sphere becomes large) and regular in the IR (when the sphere collapses to a point). Then one substitutes the solution to the supergravity action and performs holographic renormalization to obtain the partition function on gravity side. The result is supposed to match the large-$N$ limit of the localization formula of the dual field theory.

This program proved particularly fruitful for the case of $D=3$ gauge theories and their AdS${}_4$ duals. The localization formula successfully reproduces not only the $N^{3/2}$ scaling of the degrees of freedom in the strongly coupled limit, but also the coefficient which is related to the volume of the internal space \cite{Drukker:2010nc,Herzog:2010hf,Martelli:2011qj,Cheon:2011vi,Jafferis:2011zi}, for theories living on M2-branes. One also finds a nice agreement for M5-branes wrapped on 3-cycles \cite{Gang:2014qla,Gang:2014ema} which exhibit $N^3$ scaling, and D2-branes in massive IIA theory \cite{Guarino:2015jca} which exhibit $N^{5/3}$ scaling for free energy. For the gravity side analysis, we sometimes deal with, instead of 10/11 dimensional supergravity, their consistently truncated version down to four dimensions. Such theories typically contain a number of scalar fields with a potential function whose critical points provide AdS vacua. Physically speaking, non-trivial values of the scalar fields at a critical point imply that the dual field theory is at the fixed point of the renormalization group flow triggered by the field theory operators which are dual to the relevant scalar fields.

In this paper we are interested in a non-trivial supersymmetric critical point of $\cN=8, \, D=4$, $SO(8)$ gauged supergravity with $SU(3)\times U(1)$ unbroken symmetry \cite{Warner:1983vz}. This solution is 1/4-BPS, so the dual theory should be an $\cN=2$ field theory. Here the dual of the trivial vacuum is of course the celebrated ABJM theory \cite{Aharony:2008ug} with a gauge group $SU(N)\times SU(N)$ and four chiral multiplets $A_1,A_2,B_1,B_2$ in bi-fundamental representation. It is argued that the $SU(3)\times U(1)$ vacuum is dual to a fixed point one obtains after one of the chiral multiple, $A_1$ to be specific, is given a {\it superpotential} mass and integrated out \cite{Benna:2008zy}, thus deserving the name of {\it mABJM} theory \cite{Bobev:2018uxk}. According to the localization formula, the free energy of this theory on $S^3$, with R-charge assignments (also known as real masses) $\Delta_{A_2},\Delta_{B_1},\Delta_{B_2}$ to the three remaining chiral multiplets, should be 
\beq
\label{fmabjm}
F
= \frac{4\sqrt{2}\pi}{3} N^{3/2}\sqrt{\Delta_{A_2} \Delta_{B_1} \Delta_{B_2}} \,. 
\eeq

It is certainly of interest to check whether \eqref{fmabjm} can be reproduced on the gravity side as well. This problem was tackled recently in \cite{Bobev:2018wbt}, where the authors constructed the BPS equations and studied numerical solutions thereof. When one tries to perform holographic renormalization and compute the free energy, a crucial information needed is how the UV parameters (which are vevs and sources of {\it real} mass terms, according to AdS/CFT dictionary) are constrained by IR regularity. For the holographic proof of \eqref{fmabjm} according to \cite{Bobev:2018wbt}, one needs to show that $\Delta$'s from the UV expansion are {\it independent} of the IR value of a specific complex scalar, which is part of a hypermultiplet and dual to the {\it superpotential} mass term. In this paper we provide an analytic test of this statement, while \cite{Bobev:2018wbt} essentially relies on numerical solutions. 

We use a perturbative approach, which was proposed by one of us recently in \cite{Kim:2019feb}, to solve a non-conformal holography problem in this article. Calculation of the holographic free energy as a function of mass parameters and comparing the result to localization result is an interesting problem. There already exist results on $\cN=2^*$ \cite{Bobev:2013cja} and $\cN=1^*$ mass deformations \cite{Bobev:2016nua} of $\cN=4, D=4$ super Yang-Mills theory, mass deformed ABJM theory \cite{Freedman:2013ryh}, and mass-deformation \cite{Chang:2017mxc,Gutperle:2018axv} of Brandhuber-Oz theory \cite{Brandhuber:1999np,Jafferis:2012iv}. The power of our perturbative method was illustrated using representative examples in four, five, and six-dimensional AdS vacua and their mass deformations in \cite{Kim:2019feb}, and in particular we managed to obtain the holographic free energy for AdS${}_6$ problem analytically, while previously only numerical results were available in \cite{Gutperle:2018axv}. More recently we re-visited the $\cN=1^*$ problem in \cite{Kim:2019rwd} and succeeded in calculating the coefficients of the leading quartic order terms in the universal part of the holographic free energy, illustrating again the power of our perturbative prescription. 

In this article we apply our perturbative method to the BPS equations for mABJM model, constructed in \cite{Bobev:2018wbt}. There is a crucial difference here though, compared to previous works \cite{Kim:2019feb,Kim:2019rwd}. Since we have to deal with a renormalization group flow between ABJM and mABJM, the unperturbed zeroth-order solution would be a non-trivial domain wall solution in supergravity. Due to non-trivial scalar fields, explicit solutions are rare or quite complicated in general. Luckily however, the authors of \cite{Bobev:2018wbt} reported an explicit solution where the scalar fields take certain IR values, which are {\it different} from the mABJM vacuum at conformality. We choose this flow solution as a zeroth order solution for our perturbative approach. We verify the central claim of \cite{Bobev:2018wbt} on the relationship between UV parameters when we impose IR regularity, up to third order in our perturbative method. In particular, this amounts to showing that \eqref{fmabjm} indeed holds, also in holography. 

The plan of this article is as follows. Sec.2 provides a short review of mABJM theory and its gravity dual. In Sec.3 we review the BPS equations and the holographic computation done in \cite{Bobev:2018wbt}. Sec.3 is the main part where we construct perturbatively the solutions of the BPS equations. In Sec.4 we conclude with discussions. 

\section{Review of mABJM theory and its gravity dual}
The field theory description of M2-branes in  flat background is given by the ABJM model \cite{Aharony:2008ug}. It is a Chern-Simons-matter theory in $D=3$ with a quiver structure and gauge group $SU(N)\times SU(N)$. The Chern-Simons level assignment is $(k,-k)$, which leads to the orbifolding of the vacuum moduli space into $\mathbb{R}^8/\mathbb{Z}_k$. For generic integer values of $k$ the supersymmetry is $\cN=6$, while for special cases of $k=1,2$ maximal supersymmetry $\cN=8$ is restored. Readers are also reminded of the property that ABJM theory has four bi-fundamental chiral multiplets, $A_1,A_2$ in $({\bf N},{\bf \bar{N}})$ and $B_1,B_2$ in $({\bf \bar{N}},{\bf {N}})$ representation, interacting via a superpotential $W=\tfrac{4\pi}{k} \, {\rm Tr} (A_1B_1A_2B_2 - A_1B_2A_2B_1)$. 

Using supersymmetric localization \cite{Kapustin:2009kz}, one can reduce the path integral on $S^3$ to ordinary integrals. Taking the large-$N$ limit and using saddle point approximation, the free energy (logarithm of the partition function) is obtained as a function of the R-charge assignments of four bi-fundamental chiral multiplets. The result is \cite{Herzog:2010hf,Martelli:2011qj,Cheon:2011vi,Jafferis:2011zi}
\beq
F = \frac{4\sqrt{2}\pi}{3} N^{3/2}\sqrt{\Delta_{A_1} \Delta_{A_2} \Delta_{B_1} \Delta_{B_2}} \, .
\eeq
Note that, due to R-symmetry conservation of the quartic superpotential, we have a constraint
\beq
\Delta_{A_1} + \Delta_{A_2} + \Delta_{B_1} + \Delta_{B_2} = 2 \, . 
\eeq

At the true conformal point the free energy should be maximized \cite{Jafferis:2010un,Jafferis:2011zi}, giving $F = {\sqrt{2}\pi} N^{3/2}/3$ when the R-charges are all 1/2. According to AdS/CFT, this quantity is expected to match the renormalized gravitational action evaluated for $AdS_4\times S^7$, and indeed it does. On the other hand, it has been known for a long time that the maximally supersymmetric $SO(8)$-gauged supergravity in $D=4$ has a $\cN=2$ vacuum with $SU(3)\times U(1)$ supersymmetry \cite{Warner:1983vz}. Readers are referred to {\it e.g.} Table 1 in \cite{Ahn:2009as} for a list of supersymmetric and non-supersymmetric vacua. We easily see from the data there, that the ratio of the cosmological constants between the $SU(3)\times U(1)$ vacuum and the trivial vacuum is $\sqrt{27/16}$. 

Keeping this number in mind, let us now consider adding a superpotential mass term to one of the chiral multiplets (for $A_1$, to be specific) and breaking supersymmetry to $\cN=2$. Assuming that the renormalization group flow triggered by this mass deformation hits a fixed point, the R-charges should satisfy
\beq
  \Delta_{A_2} + \Delta_{B_1} + \Delta_{B_2} = 1 , \quad \Delta_{A_1} = 1 .
\eeq
Since the calculation of large-$N$ free energy apparently does not depend on superpotential, we now have \eqref{fmabjm}, whose extremized value being $F={4\sqrt{6}\pi} N^{3/2}/27$. We note that the ratio of free energy between ABJM and mABJM is again $\sqrt{27/16}$. Having the same unbroken global symmetry and free energy, it is natural to conjecture that the supergravity vacuum in question is the large-$N$ dual of mABJM theory. One can also find more non-trivial comparison between the supergravity fluctuation modes around the solution with $SU(3)\times U(1)$ symmetry and dual operators in \cite{Klebanov:2008vq,Klebanov:2009kp}.

\section{BPS equations and the holographic free energy}
\label{sec:main}
\subsection{BPS equations in conformal metric}
The lagrangian in Euclidean signature which contain the dual scalar fields of R-charge assignments and a $SU(3)$-invariant superpotential mass term is constructed in \cite{Bobev:2018wbt}. In can be expressed as follows, 
\beq
{\cal L}
& = -\frac{1}{2} R + \sum_{i=1}^3 \frac{g^{\mu \nu } \partial_\mu z_i \partial_\nu \tilde z_i}{(1-z_i \tilde z_i)^2}+\frac{g^{\mu \nu} \partial_\mu z \partial_\nu {\tilde z}}{(1-z \tilde z)^2} +\frac{1}{2L^2} {\cal P }.
\eeq
Recall that a complex scalar and its conjugate in Minkowski signature should be treated as independent when we switch to Euclidean signature. In principle they can be both complex, but for the solutions we consider here, we may consider them as real quantities. 
It is important to remember that $z_i$ and $\tilde z_i$ are {\it not} $SU(3)$ triplets. Instead, $i=1,2,3$ label three Cartan generators of $SU(4)$. On the other hand, $z,\tilde z$, from a hypermultiplet, are dual to a quadratic mass term which induces the symmetry breaking of $SU(4)$ to $SU(3)$. 

The scalar potential $\cal P$ is given in terms of superpotential ${\cal W}(z_i,z)$ and its conjugate ${\cal \widetilde W}(\tilde z_i, \tilde z)$, as follows
\beq
{\cal P} = \frac{1}{2} \left(\sum_{i=1}^3 (1-z_i \tilde z_i)^2 \nabla_{z_i}  {\cal W} \nabla_{\tilde z_i}  {\cal \widetilde W} + 4 X(1-X)^2\partial_X {\cal W} \partial_X{\cal \widetilde W}-3 {\cal W} {\cal \widetilde W}\right).
\eeq
Here the covariant derivative is defined as $\nabla_\zeta := \partial_\zeta +\frac{1}{2} \partial_\zeta (K )$, $X:= z \tilde z$. The superpotential and the K\"ahler potential are given as 
\beq
{\cal W}&= e^{K /2 } \frac{1}{1-X} (2( z_1 z_2 z_3 -1 ) + X (1-z_1)(1-z_2)(1-z_3)),
\\
e^{K /2}& = \frac{1}{(1-z_1 \tilde z_1)^{1/2}(1-z_2 \tilde z_2)^{1/2}(1-z_3 \tilde z_3)^{1/2}} .
\eeq

The theory at hand has two AdS vacua, which in M-theory setting correspond to the $\cN=8$ and a $\cN=2$ solutions respectively. At vacuum 1, all scalars vanish and ${\cal P}=-6$, which implies the radius of AdS (or hyperbolic space $\mathbb{H}^4$ to be precise) is $L$. At vacuum 2, which we call Warner vacuum, scalars take non-trivial values as follows and the supersymmetry is broken to $\cN=2$.
\beq
\label{vac2}
z_i = \tilde z_i = \sqrt{3}-2 , \quad z\tilde z = {1}/{3} \, . 
\eeq
Then the scalar potential gives ${\cal P}=-9\sqrt{3}/2$, implying the radius of AdS is now $(16/27)^{1/4}L$.

In this paper we choose conformal gauge for the metric, which is useful for our perturbative prescription as advocated in \cite{Kim:2019feb}.
\beq
\label{metric}
ds^2 = e^{2A} \left( \frac{dr^2}{r^2} + ds^2_{S^3}\right) \, . 
\eeq
The BPS equations are found from the Killing spinor equations, and are given as follows.
\beq
z_j' &= -\frac{2}{r} (1- z_j \tilde z_j )^2 \frac{ \partial_X{\cal W}}{ {\cal \widetilde W}\partial_X {\cal W} - {\cal W}  \partial_X {\cal \widetilde W }}\nabla_{\tilde z_j}  {\cal \widetilde W} ,
\\
{ \tilde z_j}' &= -\frac{2}{r} (1- z_j \tilde z_j )^2 \frac{\partial_X{\cal \widetilde W} }{ {\cal \widetilde W}\partial_X {\cal W} - {\cal W}  \partial_X {\cal \widetilde W }}  \nabla_{ z_j}  {\cal  W} ,
\\
\frac{X'}{X}  &= -\frac{8}{r} \frac{\partial_X{\cal \widetilde W} \partial_X{\cal W}}{ {\cal \widetilde W}\partial_X {\cal W} - {\cal W}  \partial_X {\cal \widetilde W }}.
\eeq
According to the derivation in \cite{Bobev:2018wbt}, the BPS conditions require $\tfrac{z}{\tilde z}$ should be constant. It is why we only have an equation for $X=z\tilde z$ here. The above equations are enough to determine all scalar fields, and they are substituted into either a differential condition 
\beq
(A')^2=\frac{1}{r^2}+\frac{1}{4r^2} e^{2A}\cal W \cal \widetilde  W , 
\eeq
or an algebraic one 
\beq
e^{2A}=\frac{16 \partial_X {\cal \widetilde W } \partial_X {\cal W}}{({\cal W}\partial_X {\cal \widetilde W }-{\cal \widetilde W }\partial_X {\cal W})^2} ,
\eeq
to determine the metric. One can check that the above equations are all consistent with each other, although it might look at first they are over-constrained.

When one turns off $z$ and $\tilde z$, we are going back to the ABJM model with general R-charge assignments. Explicit solutions are found in \cite{Freedman:2013ryh},
\beq
\label{fpsol}
z_i = c_i f(r) , \quad  \tilde z_i = {\tilde c_i} f(r) , 
\eeq
where the coefficients are related through
$c_i = {\tilde c_1}{\tilde c_2}{\tilde c_3}/{\tilde c}_i$,
and
\beq
\label{fpsol2}
f(r) = \frac{1-{\tilde c_1}{\tilde c_2}{\tilde c_3}-r^2}{1-{\tilde c_1}{\tilde c_2}{\tilde c_3}(1+r^2)} . 
\eeq
We note that this is a useful example to illustrate the power of the perturbative method \cite{Kim:2019feb}, which we employ in this paper. 
\subsection{Study of BPS solutions through UV and IR expansions}
We consider solutions which approach the trivial vacuum in the UV and become mABJM with arbitrary R-charge assignments in the IR. In terms of AdS/CFT correspondence, the R-charges of chiral multiplets in mABJM can be extracted from the leading order expansion coefficients of scalars in the following way. In the metric convention of \eqref{metric}, UV is at $r\ra 1$. In terms of the Fefferman-Graham coordinate $\rho$ which is related to $r$ as $r=1-2 e^{-\rho} + \cdots$, UV is at $\rho\ra\infty$, and the expansion for scalar fields gives that the leading terms are
\beq
z_i &= a_i e^{-\rho} + b_i e^{-2\rho} +\cdots ,  
&{\tilde z_i} &= {\tilde a_i} e^{-\rho} + {\tilde b_i} e^{-2\rho} +\cdots ,
\\
z &= a e^{-\rho} + b e^{-2\rho} +\cdots , 
&{\tilde z} &= {\tilde a} e^{-\rho} + {\tilde b} e^{-2\rho} +\cdots . 
\eeq
Then the R-charge values are given as $\Delta_{A_2} := \Delta_1,\Delta_{B_1} := \Delta_2,\Delta_{B_2} := \Delta_3$ and
\beq
\Delta_i = (a_i -{\tilde a_i})/4 . 
\eeq
One can show that the BPS equations enforce the condition $\sum_{i=1}^3 \Delta_i=1$, when $z,\tilde z$ are non-vanishing.

One finds that the above UV expansion coefficients should be related in a certain way, when one demands regularity at IR, {\it i.e.} $r=0$. It is also where the warp factor $e^{2A}$ vanishes and the sphere collapses. The IR is then characterized by the values of the scalar fields. 
\beq
c_i := z_i(0), \quad {\tilde c_i} := {\tilde z_i}(0), \quad x_0:=z(0){\tilde z}(0) \, . 
\eeq
One can show, from the analysis of BPS equations near $r=0$ \cite{Bobev:2018wbt}, 
\beq
c_i = \frac{2{\tilde c_j}{\tilde c_k}-x_0(1-{\tilde c_j})(1-{\tilde c_k})} 
{2-x_0(1-{\tilde c_j})(1-{\tilde c_k})}
,\quad
(ijk){\rm -cyclic} ,
\label{c_cons}
\eeq
and also
\beq
2({\tilde c_1}{\tilde c_2}{\tilde c_3}-1)+(1-{\tilde c_1})(1-{\tilde c_2})(1-{\tilde c_3}) = 0 \, . 
\label{tcc}
\eeq
It is clear that regular solutions are parametrized by three constants: ${\tilde c_i}$ satisfying \eqref{tcc}, and $x_0$.

In the holographic computation \cite{Bobev:2018wbt}, it is crucial to identify the UV parameters $a_i,{\tilde a_i}$ as functions of IR parameters, ${\tilde c_i},x_0$. Based on numerically constructed regular solutions, the authors of \cite{Bobev:2018wbt} conjectured that 
\beq
a_i(\tilde c,x_0) &= a^{(0)}_i({\tilde c}) + f(\tilde c,x_0) , 
\nn\\
{\tilde a_i}(\tilde c,x_0) &= {\tilde a^{(0)}_i}({\tilde c}) + f(\tilde c,x_0) . 
\eeq
In particular, it means $\Delta_i$ should be independent of $x_0$. On the other hand, $x_0=0$ is a special case where the problem reduces to pure ABJM, and exact solutions are available. It then follows that the relation for pure ABJM case, which can be derived from explicit solutions in \eqref{fpsol} and \eqref{fpsol2}, must hold more generally for $x_0\neq 0$ case as well. Summarizing, 
\beq
\Delta_i = \frac{(1+{\tilde c_j})(1+{\tilde c_k})}{(1-{\tilde c_j})(1-{\tilde c_k})} , 
\quad 
(ijk){\rm -cyclic} .
\eeq
constitutes the holographic proof of \eqref{fmabjm}. 
We confirm that this is indeed the case, using our perturbative method, in Sec.\ref{sec:4}. 
\section{Perturbative analysis}
\label{sec:4}
\subsection{Why UV should be ABJM}
Before we present a perturbative version of solutions which are ABJM at UV and mABJM in IR, let us try to answer a natural question: what if we try to construct solutions whose UV is at Warner vacuum \eqref{vac2}, and IR is mABJM with an arbitrary R-charge assignments. 

For simplicity let us consider the symmetric sector where we set $\zeta:=z_1=z_2=z_3$ and $\tilde\zeta:=\tilde{z}_1=\tilde{z}_2=\tilde{z}_3=\tilde{\zeta}$. The simplified BPS equations and the algebraic constraint can be found in the next subsection, and up to the first order we substitute
\beq
\zeta(r)&=\sqrt{3}-2+  \zeta_1 (r) \epsilon + \cdots ,
\\
\tilde \zeta(r)&=\sqrt{3}-2+ \tilde \zeta_1 (r) \epsilon +\cdots,
\\
X(r) &= {1}/{3}+ X_1 (r) \epsilon +\cdots.
\eeq
It turns out $\zeta_1,{\tilde \zeta_1}$ are given in terms of $X_1$ in the following way, 
\beq
\zeta_1(r)& =\frac{3}{8r} \left(3-2 \sqrt{3}\right) \left(1-r^2\right) X_1'(r) ,
\nn\\
\tilde \zeta_1(r)& =\frac{3r}{8} \left(3-2 \sqrt{3}\right) \left(1-r^2\right) X_1'(r) .
\eeq
And after eliminating $\zeta_1,{\tilde \zeta_1}$, we find $X_1$ should satisfy a second-order homogeneous differential equations with the following functions as two linearly independent solutions.
\beq
\left(1-r^2\right)^{\frac{1}{2}\pm \frac{\sqrt{17}}{2}}  \, _2F_1\left(\frac{1\pm \sqrt{17}}{2} ,\frac{1\pm \sqrt{17}}{2} ;1\pm \sqrt{17};1-r^2\right)
\eeq
Then it is easy to see that it is impossible to require $X_1(r=0)=0$ and make $X_1(r=1)$ finite at the same time. In more physical terms, at the Warner vacuum the superpotential mass operator cannot be treated as small. 
\subsection{Solutions for symmetric subsector}
Let us now study the symmetric sector in more detail. 
The BPS equations simplify and give
\beq
r\zeta' &=-\frac{1}{3}(1+4\zeta+\zeta^2)\frac{2(\zeta-{\tilde \zeta}^2)+X(1-\zeta) (1-{\tilde \zeta})^2}{(1-X)(\zeta-\tilde{ \zeta}) (1- \tilde{\zeta})  },
\label{zetaeq}
\\
r { \tilde \zeta}' & =-\frac{1}{3}(1+ 4\tilde{\zeta}+\tilde{\zeta}^2)\frac{2( \tilde{\zeta} - \zeta^2)+ X(1-\tilde{\zeta})(1-\zeta)^2}{(1-X)(\zeta- \tilde{\zeta}) (1-\zeta)},
\label{zetieq}
\\
 r\frac{X'}{X} & =-\frac{4}{3} \frac{(1+4\zeta+\zeta^2)(1+4 \tilde{\zeta} +\tilde{\zeta}^2)}{(\zeta-\tilde{\zeta})(1-\zeta \tilde{\zeta})},
\eeq
For the warp factor, we have 
\beq
 \left(r A'\right)^2&=1-e^{2A}\frac{ \left(X (\zeta-1)^3-2 \zeta^3+2\right) \left(X (\tilde \zeta -1)^3-2 \tilde \zeta^3+2\right)}{4  (X-1)^2 (\zeta \tilde \zeta-1)^3},
\eeq
or equivalently the algebraic constraint 
\beq
e^{2A}=-\frac{4 (\zeta^2 + 4\zeta+1) (\tilde \zeta^2 +4\tilde \zeta+1) (\zeta \tilde \zeta-1)}{9 (\zeta-1) (\tilde \zeta-1) (\zeta-\tilde \zeta)^2} . 
\eeq

Since we have autonomous equations, we may eliminate $r$ and consider only their ratios.
Explicit solutions are still not available in general, but the authors of \cite{Bobev:2018wbt} found an explicit solution. 
\beq
\label{ssol}
\tilde\zeta(\zeta)=-\zeta, \quad X(\zeta)= - \frac{4\zeta^2}{(1-\zeta^2)^2}.
\eeq
Note that we always require at UV $\zeta={\tilde \zeta}=X=0$, because we want asymptotically AdS solutions. For this particular solution at hand, at IR ${\tilde\zeta}=-{\zeta} = -2+\sqrt{3}, X=-1/3$. Requiring regularity at IR, it was found in \cite{Bobev:2018wbt} that in terms of 
$x_0\equiv X(r=0)$, general solutions should satisfy
\be
\zeta_0\equiv\zeta(r=0) = \frac{9x_0^2-12x_0+1}{(3x_0-\sqrt{3}-2)^2},
\quad
\zeti_0 \equiv \zeti(r=0) = -2 + \sqrt{3} . 
\ee

We treat this particular configuration \eqref{ssol} at $x_0=-1/3$ as our reference solution, around which we perform a perturbative analysis. For that purpose, we exploit the re-parametrization invariance and introduce a new independent variable $t$, which is related to $r$ as $dr/r=j(t)dt$. We define $t$ so that for the unperturbed solution $\zeta=-t$, and write general solutions as follows
\beq
\zeta(t)&=-t+ \sum_{n=1}^{\infty} \zeta_n (t) \epsilon^n ,
\\
\tilde \zeta(t)&=+t+ \sum_{n=1}^{\infty} \tilde \zeta_n (t) \epsilon^n ,
\\
X(t) &= \frac{-4 t^2}{(1-t^2)^2}+\sum_{n=1}^{\infty} X_n (t) \epsilon^n ,
\\
e^{2A}&=\sum_{n=0}^{\infty} w_n (t) \epsilon^n  , 
\\
j(t)&=\sum_{n=0}^{\infty}j_n (t) \epsilon^n .
\eeq

Note that now the range of the new coordinate variable $t$ is $-2+\r3\le t \le0$ at $\cO(\ep^0)$. 
Namely $t_{IR}=-2+\r3, \, t_{UV}=0$. As we consider general solutions with $\ep\neq 0$ we will keep $t_{UV}=0$ while $t_{IR}$ will receive corrections. We define in $t_0:=-2+\sqrt{3}$ for later convenience. 

Substituting the expansion into the equations, we first consider $\cO(\ep^0)$ and obtain
\beq
j_0(t)&= \frac{3 \left(t^2+1\right)^2}{\left(1-t^2\right) \left(t^4-14 t^2+1\right)} , 
\\
w_0(t)&= \frac{(t^2+1)(t^4-14t^2+1)}{9 t^2(1-t^2)}.
\eeq
Note that the warp factor is now called $w$ and it satisfies $w(t_{UV})=\infty$ and $w(t_{IR})=0$, as it should. 

As we turn to higher orders, we fix the re-parametrization gauge and set $X_n(t)=0, \, n\ge 1$. Then at first order we discover $j_1,w_1$ are given as certain linear combinations of $\zeta_1,\tzeta_1$. We also have a coupled first-order differential equations for $\zeta_1,\tzeta_1$, which can be explicitly solved 
\beq
\zeta_1(t)&=\frac{\left(\sqrt{3}+3\right) t (1+t)}{3 \left(\left(\sqrt{3}-2\right) t-1\right)}, 
\\
\tilde \zeta_1(t)&= \frac{\left(\sqrt{3}+3\right) t (1-t) }{3 \left(t+\sqrt{3}-2\right)}.
\eeq
Note that the UV boundary condition $\zeta(t_{UV})=\tzeta(t_{UV})=0$ is satisfied. Substituting the above result into the expressions for $j_1,w_1$, we obtain 
\beq
j_1 &= \frac{\left(2 \sqrt{3}+3\right) \left(t^2+1\right)^2 \left(t^2+6 \sqrt{3} t-1\right)}{\left(t^4-14 t^2+1\right)^2},
\\
w_1 &= \frac{2\left(3+2\r3\right) \left(t^2+1\right) \left( t^4 + 3\r3 t^3 - 8t^2 - 3\r3 t +1 \right)}{27  t^2 \left(t^2-1\right)} . 
\eeq

We can push explicit integration to next order $\cO(\ep^2)$. Again $\zeta_2,\tzeta_2$ satisfy coupled first-order differential equations. The homogeneous part is the same as the one for $\zeta_2,\tzeta_2$, but this time we have additionally an inhomogeneous part expressed in terms of $\zeta_1,\tzeta_1$. Solutions with the right UV boundary condition are given as follows
\beq
\zeta_2(t)
&=- \frac{(2+\sqrt{3})t (t+1) \left(\left(19 \sqrt{3}+33\right) t^2+\left(4 \sqrt{3}+4\right) t-13 \sqrt{3}-23\right)}{6  \left(t+\sqrt{3}-2\right) \left(t+\sqrt{3}+2\right)^2}
\nn\\
&-\frac{\left(7+4\sqrt{3}\right)  t (1-t^2) }{3  \left(t-\sqrt{3}+2\right) \left(t+\sqrt{3}+2\right)}\log \left[ \frac{2(1-t^2)}{1-2 \sqrt{3} t-t^2}\right],
\\
\tilde \zeta_2(t)
&=
\frac{(3\sqrt{3}+5)\left(3 t-\sqrt{3}-2\right) \left(t-\sqrt{3}+2\right) t (1-t) }{6 \left(t+\sqrt{3}-2\right)^2 \left(t+\sqrt{3}+2\right)}
\nn\\
&+\frac{\left(7+4\sqrt{3}\right) t (1-t^2) }{3  \left(t-\sqrt{3}-2\right) \left(t+\sqrt{3}-2\right)}  \log \left[ \frac{2(1-t^2)}{1-2 \sqrt{3} t-t^2}\right]. 
\eeq
One might be alarmed by the logarithms in the result, which is usually a signal for conformal anomaly in even dimensions. However, the logarithms above and in higher order results do not cause log terms in UV $t=0$ obviously, and has nothing to do with conformal anomaly. On the determination of $\zeta_2,\tilde \zeta_2$,
to be more specific one of the integration constants are fixed by regularity of $\zeta_2$ at IR, and the other constant is fixed by our choice $\tilde\zeta_2(t_0)=0$.
It is then straightforward to write down the next-order functions $\zeta_3,\zeti_3$ in an integral form. Because the result is quite lengthy with a lot of polylogarithms, we will not present the $\cO(\ep^3)$ solutions explicitly here. Knowing $\zeta_3',\zeti_3'$ explicitly is nonetheless useful in the following analysis.

Now let us try to extract the crucial information on this holographic problem. The IR value of $t$ should be determined through the requirement $\zeti(t=t_{IR})=-2+\r3$. There should be a series expansion for $t_{IR}$ in terms of $\ep$. It turns out
\be
t_{IR}= \sqrt{3}-2-\epsilon-\left(\frac{3 }{2}+\frac{2 }{\sqrt{3}}\right)\epsilon ^2-\left(\frac{11 }{6}+\sqrt{3}\right) \epsilon ^3+{\cal O}(\epsilon^4) .
\ee
Inverting it and recalling $X(t_{IR})=x_0$, we have 
\be
x_0= -\frac{1}{3}-\frac{4}{9} \left(3+2 \sqrt{3}\right) \epsilon-\left(\frac{73}{9}+\frac{14}{\sqrt{3}}\right) \epsilon ^2-\left(\frac{317}{9}+\frac{61}{\sqrt{3}}\right) \epsilon ^3+{\cal O}(\epsilon^4) .
\ee

In order to determine $a_s:=a_1=a_2=a_3,{\tilde a}_s:={\tilde a_1}={\tilde a_2}={\tilde a_3}$, and $f(x_0)$, we need to analyze the UV (near $t=0$) behavior. Let us first confirm that $a_s-{\tilde a}_s=4/3$: namely, it is independent of $\ep$. In \cite{Bobev:2018wbt} it was derived from the UV expansion of the BPS equations, so for us it is a quick consistency test. This can be seen without knowing the perturbative results $w_n,\zeta_n,\zeti_n$ explicitly. Note that $a_s,{\tilde a}$ are just given by the UV limits (near $t=0$) of the right-hand-side expressions in Eq.{\eqref{zetaeq} and Eq.\eqref{zetieq}}. Since $\zeta,\zeti \sim \cO(t)$ and $X\sim \cO(t^2)$, we easily see that
\be
a_s = \lim_{t\ra 0} \frac{4\zeta}{3(\zeta-\zeti)} , \quad
{\tilde a}_s = \lim_{t\ra 0} \frac{4\zeti}{3(\zeta-\zeti)} . 
\ee
Then obviously $a_s-{\tilde a}_s = 4/3$, as desired. This argument can be easily generalized to non-symmetric case. 

Calculating $f(x_0)\equiv a_s - 2/3 + 2\r3/9$ requires doing the integration for $\zeta_3,\zeti_3$ explicitly, which we did not manage to accomplish. Using numerical integration, we find 
\footnote{It is tempting to conjecture 0.62107... divided by $\sqrt{3}$, $1.7573$, is a rational number, but it seems unlikely.}
\beq
f(x_0)&=\tfrac{2\r3}{9}-\tfrac{\sqrt{3}}{2}  \left(x_0+1/3\right)-\tfrac{9\sqrt{3} }{32} \left(x_0+1/3\right)^2
\nn\\
&- 0.621072954165398 \left(x_0+1/3\right)^3+{\cal O}\left((x_0+1/3)^4\right).
\label{fx0}
\eeq
Figure 1 is a plot of the truncated cubic expression from the above expression. When we compare it with Figure 4 of \cite{Bobev:2018wbt}, we find a reasonably good agreement roughly in the range of $-0.45\lesssim r \lesssim 0.25$. 
As a consistency check we may evaluate $f(0)$, which should be zero, from the above expression. We obtain $0.019$ instead.
\begin{figure}[h]
\centering
  \includegraphics[width=0.5\linewidth]{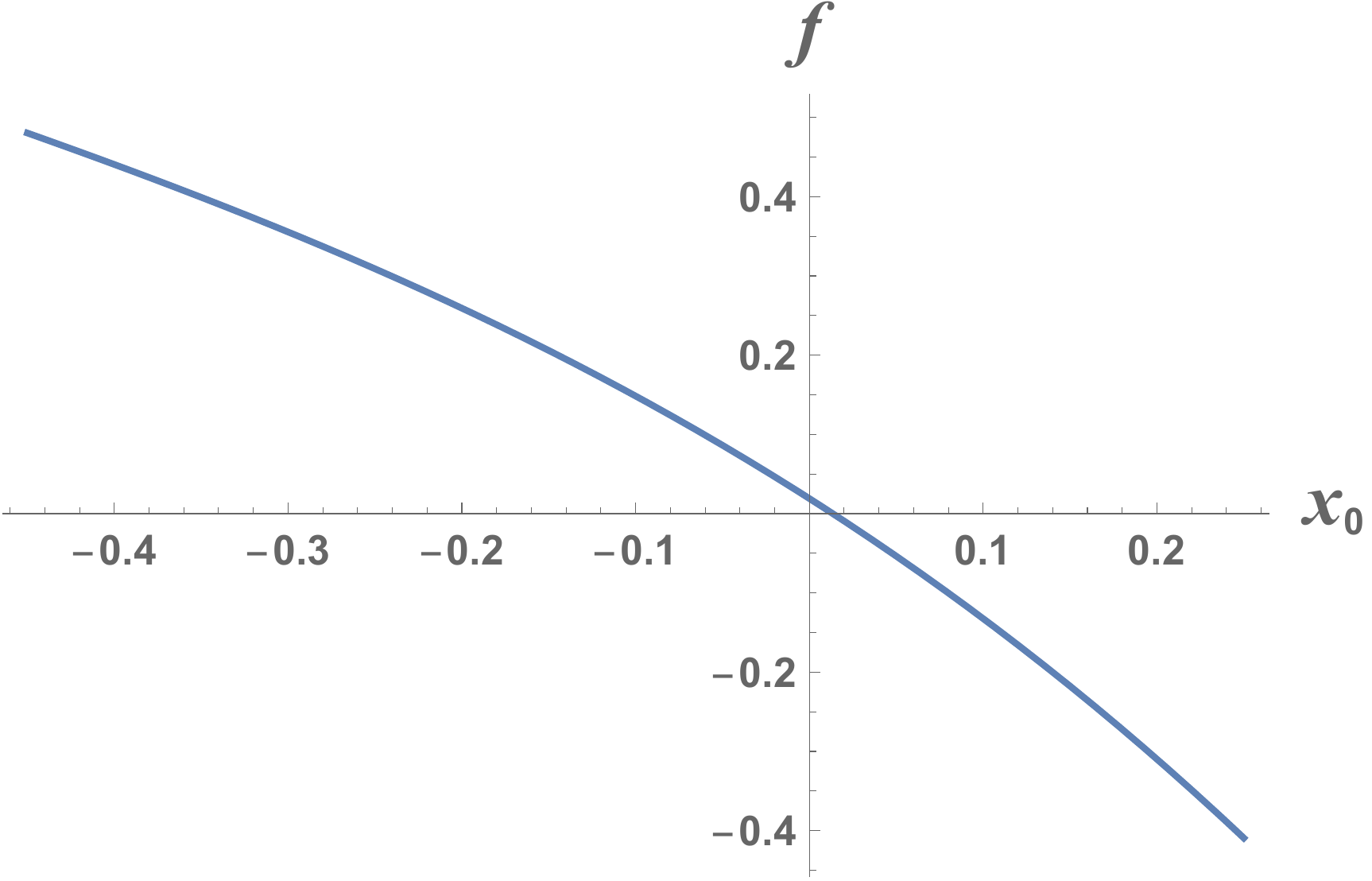}
  \caption{Plot of $f(x_0)$ using Eq.\eqref{fx0}, truncated to cubic order in $(x_0+1/3)$.}
\end{figure}
\subsection{Non-symmetric solutions}
We use basically the same strategy also to construct general solutions. In particular we use a new holographic coordinate $t$ defined through $dr/r=j(t)dt$, and using the gauge freedom fix $X(t)= -4t^2/(1-t^2)^2$. 
Our $\ep$ expansion is now
\beq
z_i(t)=-t+ \sum_{n=1}^{\infty} z_{i,n} (t) \epsilon^n ,
\quad
{\tilde z}_i(t)=+t+ \sum_{n=1}^{\infty} {\tilde z}_{i,n} (t) \epsilon^n ,
\eeq
and
\beq
e^{2A}=\sum_{n=0}^{\infty} W_n (t) \epsilon^n  , 
\quad
j(t)=\sum_{n=0}^{\infty}J_n (t) \epsilon^n .
\eeq

It is obvious that $J_0(t)=j_0(t)$ and $W_0(t)=w_0(t)$: the same as the symmetric solution. At $\cO(\ep)$ the solutions are essentially the same as the symmetric case and all $z_{i,1}\, ({\tilde z}_{i,1})$ have the same profile. Namely, we have
\beq
z_{i,1} = \ep_i \zeta_1 (t) , \quad {\tilde z}_{i,1} = \ep_i \zeti_1 (t) . 
\eeq
We now absorb $\ep$ into $\ep_i$, which are treated as small expansion parameters. Introducing three parameters here is obviously consistent with the fact that general solutions are parametrized by ${\tilde c_i},x_0$. In terms of our perturbative solutions, ${\tilde c_i},x_0$ are functions of $\ep_i$.
For the rest of $\cO(\ep)$ functions,
\beq
J_1 (t) = \frac{\ep_1+\ep_2+\ep_3}{3} j_1(t) , \quad W_1(t) =  \frac{\ep_1+\ep_2+\ep_3}{3} w_1(t) . 
\eeq

At the next order $\cO(\ep^2)$, we find due to symmetry reasons the solutions should take the following form:
\beq
z_{i,2} &= \ep_i^2 A(t) + \ep_i (\ep_1+\ep_2+\ep_3) B(t) + \frac{\ep_1\ep_2\ep_3}{\ep_i} C(t) , 
\\
{\tilde z}_{i,2} &= \ep_i^2 {\tilde A}(t) + \ep_i (\ep_1+\ep_2+\ep_3) {\tilde B}(t) + \frac{\ep_1\ep_2\ep_3}{\ep_i} {\tilde C}(t) . 
\eeq
The component functions are found to be 
\beq
A &=-\frac{\left(87 \sqrt{3}+151\right) t (t+1) \left(47 t^2+51 \sqrt{3} t-48 t+10 \sqrt{3}-26\right)}{423 \left(t+\sqrt{3}-2\right) \left(t+\sqrt{3}+2\right)^2}
   \nn\\
  & -\frac{\left(7+4 \sqrt{3}\right)
   t(t^2-1)  }{9
   \left(t-\sqrt{3}+2\right)
   \left(t+\sqrt{3}+2\right)}\log \left[\frac{2
   \left(1-t^2\right)}{1-2
   \sqrt{3} t-t^2}\right],
   \\
B&= -\frac{\left(11 \sqrt{3}+19\right) t (t+1) \left(t^2-2 \sqrt{3} t-1\right)}{18 \left(t+\sqrt{3}-2\right) \left(t+\sqrt{3}+2\right)^2} 
   \nn\\
& +\frac{\left(7+4 \sqrt{3}\right)
   t(t^2-1)  }{9
   \left(t-\sqrt{3}+2\right)
   \left(t+\sqrt{3}+2\right)}   \log \left[\frac{2
   \left(1-t^2\right)}{1-2
   \sqrt{3} t-t^2}\right] ,
   \\
C&=- \frac{\left(3 \sqrt{3}+5\right) \left(t-\sqrt{3}-2\right) t (t+1)}{9 \left(t+\sqrt{3}-2\right) \left(t+\sqrt{3}+2\right)}
   \nn\\
&  +\frac{\left(7+4 \sqrt{3}\right)
   t (t^2-1) }{9
   \left(t-\sqrt{3}+2\right)
   \left(t+\sqrt{3}+2\right)} \log \left[\frac{2
   \left(1-t^2\right)}{1-2
   \sqrt{3} t-t^2}\right],
\eeq
and
\beq
\tilde A & =-\frac{(t-1) t \left(2 t^2-3 \sqrt{3} t+3 t-\sqrt{3}+1\right)}{9 \left(t+\sqrt{3}-2\right)^2 \left(\sqrt{3} t-2 t-1\right)}
   \nn\\
   & +\frac{
   \left(7+4 \sqrt{3}\right) t (t^2-1) 
    }{9
   \left(t-\sqrt{3}-2\right)
   \left(t+\sqrt{3}-2\right)}
   \log \left[\frac{2
   \left(1-t^2\right)}{1-2
   \sqrt{3} t-t^2}\right],
\\
\tilde B & =-\frac{\left(3 \sqrt{3}+5\right) (t-1) t \left(t^2-2 \sqrt{3} t-1\right)}{18 \left(t+\sqrt{3}-2\right)^2 \left(t+\sqrt{3}+2\right)}
   \nn\\
   &
  - \frac{\left(7+4 \sqrt{3}\right)
t (t^2-1) }{9
   \left(t-\sqrt{3}-2\right)
   \left(t+\sqrt{3}-2\right)} \log \left[\frac{2
   \left(1-t^2\right)}{1-2
   \sqrt{3} t-t^2}\right],
\\ 
\tilde C & =-\frac{\left(11 \sqrt{3}+19\right) \left(t-\sqrt{3}+2\right) (t-1) t}{9 \left(t+\sqrt{3}-2\right) \left(t+\sqrt{3}+2\right)}
   \nn\\
   & - 
   \frac{\left(7+4 \sqrt{3}\right)
    t (t^2-1) }{9
   \left(t-\sqrt{3}-2\right)
   \left(t+\sqrt{3}-2\right)} \log \left[\frac{2
   \left(1-t^2\right)}{1-2
   \sqrt{3} t-t^2}\right].
\eeq
It is also straightforward to calculate the warp factor at $\cO(\ep^2)$,
\beq
W_2(t)&=-\frac{\left(5+3 \sqrt{3}\right) \left(t^2+1\right) \left(2 \left(t^4-\left(7-\sqrt{3}\right) t^2+1\right)+9 \left(1+\sqrt{3}\right) t \left(t^2-1\right)\right)}{162 t^2 \left(t^2-1\right)}\left(\epsilon_1^2+ \epsilon_2^2+\epsilon_3^2\right)
\nn\\
&-\frac{2 \left(7+4 \sqrt{3}\right) \left( t^2+1\right) \left(t^6+6 \sqrt{3} t^5+11 t^4-8 \sqrt{3} t^3-11 t^2+6 \sqrt{3} t-1\right)}{81 t^2 \left(t^2-1\right) \left(t^2+2 \sqrt{3} t-1\right)}\left(\epsilon_1 \epsilon_2+\epsilon_3 \epsilon_2+\epsilon_1 \epsilon_3\right)
\nn\\
&+\frac{2 \left(7+ 4 \sqrt{3}\right)  \left(t^2+1\right) \left(t^4-8 t^2+1\right)}{81  t^2 (1-t^2)}\log \left[\frac{2(1-t^2)}{1-2 \sqrt{3} t-t^2}\right]\left(\epsilon_1 \epsilon_2+\epsilon_3 \epsilon_2+\epsilon_1 \epsilon_3\right).
\eeq
Since $t_{IR}$ is where the warp factor $e^{2A}$ vanishes, one obtains
\beq
t_{IR}&=\sqrt{3}-2-\frac{1}{3} \left(\epsilon_1+\epsilon_2+\epsilon_3\right) -\frac{1}{108} \left(15+8 \sqrt{3}\right) \left(\epsilon_1+\epsilon_2+\epsilon_3\right){}^2-\frac{1}{12} \left(\epsilon_1^2+\epsilon_2^2+\epsilon_3^2\right) 
\nn\\
&-\frac{1}{324} \left(23+13 \sqrt{3}\right) \left(\epsilon_1+\epsilon_2+\epsilon_3\right){}^3+\frac{1}{27} \left(9+5 \sqrt{3}\right) \epsilon_1 \epsilon_2 \epsilon_3
\nn\\
&-\frac{1}{324} \left(27+11 \sqrt{3}\right) \left(\epsilon_1^3+\epsilon_2^3+\epsilon_3^3\right)+{\cal O} (\epsilon^4).
\eeq
Now we can calculate the IR values of the fields by substituting $t_{IR}$ above into the perturbative solutions of scalar fields. First the hypermultiplet scalar $X$ at IR is
\beq
x_0&=-\frac{1}{3}-\frac{4}{27} \left(3+2 \sqrt{3}\right) \left(\epsilon_1+\epsilon_2+\epsilon_3\right)-\frac{10}{81}  \left(7+4 \sqrt{3}\right) \left(\epsilon_1+\epsilon_2+\epsilon_3\right){}^2
\nn\\
&-\frac{1}{27} \left(3+2 \sqrt{3}\right) \left(\epsilon_1^2+\epsilon_2^2+\epsilon_3^2\right) 
-\frac{2}{729}  \left(480+277 \sqrt{3}\right) \left(\epsilon_1+\epsilon_2+\epsilon_3\right){}^3
\nn\\
&+\frac{1}{81} \left(115+66 \sqrt{3}\right) \epsilon_1 \epsilon_2 \epsilon_3-\frac{1}{243} \left(88+51 \sqrt{3}\right) \left(\epsilon_1^3+\epsilon_2^3+\epsilon_3^3\right)+{\cal O} (\epsilon^4).
\eeq
And the vector multiplet scalars $z_i, \tilde z_i$ give
\beq
c_i&=-\sqrt{3}+2+\frac{1}{3}\epsilon _i+\frac{1}{3} \left(\epsilon _1+\epsilon _2+\epsilon _3\right)+\frac{1}{18}\left(3+4 \sqrt{3}\right) \frac{ \epsilon _1 \epsilon _2 \epsilon _3}{\epsilon _i}+\frac{5 }{18}\epsilon _i^2+\frac{1}{27} \left(3+4 \sqrt{3}\right) \epsilon _i \left(\epsilon _1+\epsilon _2+\epsilon _3\right)
\nn\\
&+\frac{2}{27} \left(3+\sqrt{3}\right) \left(\epsilon _1+\epsilon _2+\epsilon _3\right){}^2+\frac{1}{108} \left(19+8 \sqrt{3}\right) \epsilon _i^3-\frac{1}{972} \left(62+53 \sqrt{3}\right) \left(\epsilon _1^3+\epsilon _2^3+\epsilon _3^3\right)
\nn\\
&+\frac{5}{324} \left(21+13 \sqrt{3}\right) \epsilon _i^2 \left(\epsilon _1+\epsilon _2+\epsilon _3\right)-\frac{1}{108} \left(20+13 \sqrt{3}\right) \epsilon _i \left(\epsilon _1+\epsilon _2+\epsilon _3\right){}^2
\nn\\
&+\frac{1}{972} \left(212+125 \sqrt{3}\right) \left(\epsilon _1+\epsilon _2+\epsilon _3\right){}^3-\frac{1}{162} \left(41+20 \sqrt{3}\right) \epsilon _1 \epsilon _2 \epsilon _3+{\cal O} (\epsilon^4),
\nn\\
\tilde c_i&=\sqrt{3}-2+\epsilon _i-\frac{1}{3} \left(\epsilon _1+\epsilon _2+\epsilon _3\right)+\frac{1}{6}\frac{\epsilon _1\epsilon _2 \epsilon _3}{\epsilon _i}-\frac{1}{6}\epsilon _i^2+\frac{2}{9} \left(3+\sqrt{3}\right) \epsilon _i\left(\epsilon _1+\epsilon _2+\epsilon _3\right) 
\nn\\
&-\frac{8}{108} \left(3+\sqrt{3}\right) \left(\epsilon _1+\epsilon _2+\epsilon _3\right){}^2+\frac{1}{36} \left(9+4 \sqrt{3}\right) \epsilon _i^3-\frac{1}{324} \left(20+7 \sqrt{3}\right) \left(\epsilon _1^3+\epsilon _2^3+\epsilon _3^3\right)
\nn\\
&-\frac{1}{36} \left(11+5 \sqrt{3}\right) \epsilon _i^2 \left(\epsilon _1+\epsilon _2+\epsilon _3\right)+\frac{1}{108} \left(40+21 \sqrt{3}\right) \epsilon _i \left(\epsilon _1+\epsilon _2+\epsilon _3\right){}^2
\nn\\
&-\frac{1}{324} \left(30+17 \sqrt{3}\right) \left(\epsilon _1+\epsilon _2+\epsilon _3\right){}^3+\frac{1}{54} \left(1+2 \sqrt{3}\right) \epsilon _1 \epsilon _2 \epsilon _3+{\cal O} (\epsilon^4).
\eeq
We have verified the constraints \eqref{c_cons} and \eqref{tcc} are indeed satisfied by our results. 

We can also calculate the UV parameters, and in particular we need $a_i,{\tilde a_i}$ in order to
calculate $\Delta_i = (a_i-\tilde a_i)/4$. It turns out that this time we were only able to evaluate some of cubic order coefficients numerically, and 
\beq
a_i(\epsilon)&=\frac{2}{3}+\frac{2}{27} \left(3\left(3 +\sqrt{3}\right) \epsilon_i+\left(3+2 \sqrt{3}\right) \left(\epsilon_1+\epsilon_2+\epsilon_3\right)\right) 
\nn\\
&+\frac{4}{81} \left(5+2 \sqrt{3}\right) \epsilon_i^2+\frac{1}{81} \left(8+5 \sqrt{3}\right) \left(\epsilon_1^2+\epsilon_2^2+\epsilon_3^2\right)
\nn\\
&+\frac{1}{81} \left(73+43 \sqrt{3}\right)  \epsilon_i \left(\epsilon_1+\epsilon_2+\epsilon_3\right)+\frac{2 }{81}\left(71+41 \sqrt{3}\right)\frac{ \epsilon_1 \epsilon_2 \epsilon_3}{\epsilon_i}
\nn\\
&+1.35876 \epsilon_i^3-1.8564 \left(\epsilon_1^3+\epsilon_2^3+\epsilon_3^3\right)+5.47781\epsilon_i^2 \left(\epsilon_1+\epsilon_2+\epsilon_3\right) 
\nn\\
&-3.58893 \epsilon_i \left(\epsilon_1+\epsilon_2+\epsilon_3\right){}^2+1.7595\left(\epsilon_1+\epsilon_2+\epsilon_3\right){}^3 +2.97386 \epsilon_1 \epsilon_2 \epsilon_3 +{\cal O} (\epsilon^4),
\\
\tilde a_i(\epsilon)&=-\frac{2}{3}+\frac{2}{27} \left(3\left(9+5 \sqrt{3}\right) \epsilon_i-\left(3+2 \sqrt{3}\right) \left(\epsilon_1+\epsilon_2+\epsilon_3\right)\right) 
\nn\\
& -\frac{2}{81}  \left(1+\sqrt{3}\right) \epsilon_i^2-\frac{1}{81} \left(8+5 \sqrt{3}\right) \left(\epsilon_1^2+\epsilon_2^2+\epsilon_3^2\right)
\nn\\
&+\frac{1}{81} \left(143+83 \sqrt{3}\right) \left(\epsilon_1+\epsilon_2+\epsilon_3\right) \epsilon_i+\frac{2}{81}\left(1+\sqrt{3}\right) \frac{\epsilon_1 \epsilon_2 \epsilon_3}{ \epsilon_i}
\nn\\
&+2.62989\epsilon_i^3 -0.387987 \left(\epsilon_1^3+\epsilon_2^3+\epsilon_3^3\right)-3.03674\epsilon_i^2 \left(\epsilon_1+\epsilon_2+\epsilon_3\right) 
\nn\\
&+3.07307 \epsilon_i \left(\epsilon_1+\epsilon_2+\epsilon_3\right){}^2+0.484888\left(\epsilon_1+\epsilon_2+\epsilon_3\right){}^3 -2.7025\epsilon_1 \epsilon_2 \epsilon_3 +{\cal O} (\epsilon^4).
\eeq
We are now ready to check the central claim in \cite{Bobev:2018wbt}, {\it i.e.} for $(ijk)$-cyclic,
\beq
a_i (\epsilon) = \frac{4\tilde c_j \tilde c_k}{1- \tilde c_1 \tilde c_2 \tilde c_3} + f(x_0,\tilde c) ,
\quad
\tilde a_i (\epsilon)  = \frac{ 4\tilde c_i }{ 1- \tilde c_1 \tilde c_2 \tilde c_3}  + f(x_0,\tilde c) .
\eeq
and 
\beq
 f(x_0,\tilde c) &= \frac{2\sqrt{3}}{9}   -\frac{9\sqrt{3}}{32}(x_0+1/3)^2-0.621963(x_0+1/3)^3 \nn\\
 & -\prod_{i=1}^3\left(\tilde c_i-\sqrt{3}+2\right) 
 \left[ 0.480881 
 +0.200891 (x_0 + 1/3) \sum_{i=1}^3 (\tilde c_i-\sqrt{3}+2 )^{-1} 
 \right]
 \nn\\
& +{\cal O} (\epsilon^4).
\eeq
Let us comment here that the above expression reduces to \eqref{fx0} when $\ep_1=\ep_2=\ep_3$ as it should, and the above formula is in fact independent of ${\tilde c}_i$ up to ${\cal O}(\ep^2)$, since the second line is $\cO(\ep^3)$.

\section{Discussions}
In this paper we applied our perturbative prescription \cite{Kim:2019feb,Kim:2019rwd} to mABJM theory and confirmed the conjecture of \cite{Bobev:2018wbt}. The analysis reported here is a rather non-trivial extension of \cite{Kim:2019feb,Kim:2019rwd}, where the unperturbed solutions were pure AdS, while we perturb around a non-trivial flow solution which connects two distinct vacua. We have illustrated that our method is still effective and have provided analytic confirmation of the holographic free energy formula \eqref{fmabjm}. We plan to tackle other problems with our method to obtain analytic expression for the free energy of other quantities accessible via localization technique and holography.

\section*{Acknowledgement}
We thank Hyojoong Kim for useful discussions, and in particular for drawing our attention to \cite{Bobev:2018wbt} when it first appeared on the arXiv. 
This work was completed during a visit to CERN, and we thank them for hospitality.
This research was supported by the National Research Foundation of Korea (NRF) grant 2018R1D1A1B07045414. 
Se-Jin Kim was also partly supported by the CERN-Korea graduate student visiting program supported by NRF.



\bibliographystyle{elsarticle-num} 
\bibliography{mabjm}





\end{document}